\documentclass[utf8]{FrontiersinHarvard} 

\usepackage{url,hyperref,lineno,microtype,subcaption}
\usepackage[onehalfspacing]{setspace}
\usepackage{graphicx}        
\usepackage{amssymb}        
\usepackage{color}           
\usepackage{breakurl}        
\usepackage{tabularx}
\usepackage{multirow}
\usepackage{wasysym}

\usepackage{multirow}
\usepackage{amsmath}
\usepackage{amsfonts}

\newcommand\at[2]{\left.#1\right|_{#2}}

\def\keyFont{\fontsize{8}{11}\helveticabold }
\def\firstAuthorLast{Zavershinskii {et~al.}} 
\def\Authors{Dmitrii I. Zavershinskii\,$^{1,2,*}$, Nonna E. Molevich\,$^{1,2}$, Dmitrii S. Riashchikov\,$^{1,2}$ and Sergey A. Belov\,$^{1,2}$}
\def\Address{$^{1}$ Department of Physics, Samara National Research University, 34 Moscovskoe sh.,  Samara,  443086, Russia \\
$^{2}$ Department of Theoretical Physics, Lebedev Physical Institute, 221 Novo-Sadovaya st., Samara, 443011, Russia }
\def\corrAuthor{Dmitrii I. Zavershinskii}
\def\corrEmail{d.zavershinskii@gmail.com}

\begin{document}
\onecolumn
\firstpage{1}

\title[Exact description of waves in coronal loops]{Exact solution to the problem of slow oscillations in coronal loops and its diagnostic applications} 

\author[\firstAuthorLast ]{\Authors} 
\address{} 
\correspondance{} 

\extraAuth{}

\maketitle

\begin{abstract}

Magnetoacoustic oscillations are nowadays routinely observed in various regions of the solar corona. This allows them to be used as means of diagnosing plasma parameters and processes occurring in it. Plasma diagnostics, in turn, requires a sufficiently reliable MHD model to describe the wave evolution.  In our paper, we focus on obtaining the exact analytical solution to the problem of the  linear evolution of standing slow magnetoacoustic (MA) waves in coronal loops. Our consideration of the properties of slow waves is conducted using the infinite magnetic field assumption.  The main contribution to the wave dynamics in this assumption comes from such processes as thermal conduction, unspecified coronal heating, and optically thin radiation cooling.  In our consideration, the wave periods are assumed to be short enough so that the thermal misbalance has a weak effect on them. Thus, the main non-adiabatic process affecting the wave dynamics remains thermal conduction. The exact solution of the evolutionary equation is obtained using the Fourier method. This means that it is possible to trace the evolution of any harmonic of the initial perturbation, regardless of whether it belongs to entropy or slow mode.  We show that the fraction of energy between entropy and slow mode is defined by the thermal conduction and coronal loop parameters. It is shown for which parameters of coronal loops it is reasonable to associate the full solution with a slow wave, and when it is necessary to take into account the entropy wave. Furthermore, we obtain the relationships for the phase shifts  of various plasma parameters applicable to any values of harmonic number and thermal condition coefficient. In particular, it is shown that the phase shifts between density and temperature perturbations for the second harmonic of the slow wave vary between $\pi/2$ to 0, but are larger than for the fundamental harmonic. The obtained exact analytical solution could be further applied to the interpretation of observations and results of numerical modelling of slow MA waves in the corona.

\tiny
 \keyFont{ \section{Keywords:} Solar corona, magnetoacoustic wave, slow wave, entropy wave, MHD, Sun, exact solution, thermal conduction} 
\end{abstract}

\section{Introduction}
\label{s:Introduction} 

Nowadays, the processes in the solar atmosphere are actively studied with spaceborne and ground-based instruments. The accuracy of instruments has made it possible to detect signals from individual structural elements of the solar atmosphere like coronal loops, spicules, etc. The observed periodic or quasi-periodic signals can be readily associated with the magnetoacoustic (MA) waves evolving in these structures (see \cite{2020ARA&A..58..441N} for a recent review). Developed MHD theories and amount of observed events allow us to use MA waves as a diagnostic tool for the coronal plasma, giving birth to such a research field as coronal seismology. Speaking of some applications, the observations of fast MA waves are used for estimations of the density stratification \citep{2005ApJ...624L..57A}, corona loop parameters \citep{2015ApJ...812...22C}, etc. For the most recent reviews, we refer the reader to the works of \cite{2021SSRv..217...73N, 2020SSRv..216..136L, 2021SSRv..217...76B}. In turn, the slow MA waves are used to seismologically infer coronal temperature  \citep{2009ApJ...706L..76M}, transport coefficients \citep{2015ApJ...811L..13W}, adiabatic index \citep{2018ApJ...868..149K},  magnetic field strength and local Alfvens speed \citep{2016NatPh..12..179J,2017ApJ...837L..11C}, or even to introduce some constraints on enigmatic coronal heating function \citep{2020A&A...644A..33K}. Prospects for the application of slow waves for coronal plasma diagnostics are discussed in detail in a recent review by \cite{2021SSRv..217...34W}.  The diagnostics of plasma parameters requires a sufficiently reliable MHD  model and accurate observational data analysis. In this paper, we will be focused on the former, i.e. developing an advanced exact analytical model of slow waves in the corona.

The use of MA waves as a seismological tool involves taking into account the primary processes, which determine the dispersion properties of waves. For magnetoacoustic waves (fast and slow), it is important to correctly take into account the influence of magnetic structuring on wave properties. For such needs, one can apply the classical approach from pioneering works by \citet{1975IGAFS..37....3Z, 1982SvAL....8..132Z} and \citet{1983SoPh...88..179E}, representing the coronal loop as a straight plasma cylinder. Despite the generality of the given description, this approach is quite complicated even under the assumption of ideal adiabatic plasma. In this regard, to describe MA  waves, one often uses the thin flux tube approximation \citep{Zhugzhda96}. This approach allows us to easily analytically show the wave-guiding dispersion of slow waves, which, in particular, is manifested as the variation of the slow-wave phase speed from the usual sound speed for short periods to the so-called tube speed in the long period limit. The latter speed is frequently used for the estimation of loop magnetic field \citep{2007ApJ...656..598W, 2016NatPh..12..179J} and other loop parameters.  However, if the plasma $\beta$ is sufficiently low, the mentioned dispersion effect on slow waves becomes rather weak. In this case, the modelling can be further simplified for the slow waves propagating along the magnetic field lines. This approach is known as an infinite magnetic field approximation (see e.g. \cite{Zavershinskii2019},  \cite{2019A&A...628A.133K}, \cite{2018ApJ...856...51R} for details). 

Thus, one may show that in a highly magnetized plasma, the main dispersion sources for slow waves become thermal conduction and non-adiabatic processes. It has been shown in \cite{Zavershinskii2019, 2019A&A...628A.133K}, using the infinite magnetic field approximation, that the long-period limit of the slow-wave phase velocity is highly affected by the coronal heating and cooling rates. Moreover, this statement is valid for any plasma $\beta$ less than unity. It has been confirmed within the thin flux tube approximation \citep{2021SoPh..296..122B} and using slab geometry \citep{2022MNRAS.514.5941A}. Additionally, plasma heating/cooling can lead to the amplification/attenuation of slow waves depending on the form of non-adiabatic functions. The mentioned damping can support the dissipation by thermal conduction and can be used to introduce constraints for the coronal heating function \citep{2020A&A...644A..33K, 2023arXiv230109012K}. In turn, the amplification by non-adiabatic processes may lead to suppressed damping or even the growth of slow waves. Combination of dispersion effects and amplification can lead to the formation of quasi-periodic patterns (see \cite{Zavershinskii2019} for details). If the amplification is not balanced by the thermal conduction damping, then it will be limited by non-linear processes allowing autowave-shock structures to exist \citep{2010PhPl...17c2107C, 2020PhRvE.101d3204Z, 2021PhFl...33g6110M}. In addition, non-adiabatic processes introduce the phase-shift between perturbation of different parameters (e.g. temperature and density), which also depends on the form of heat-loss functions \citep{2021SoPh..296...96Z}. The problem of heating/cooling influence on the phase shift of propagating slow waves has been investigated by  \cite{2021SoPh..296...20P, 2022BLPI...49..282M} and by  \cite{2022SoPh..297....5P} for standing modes. It should be noted, that heating/cooling affects not only slow waves but also entropy waves \citep{2007AstL...33..309S}. The latter waves can be amplified separately, with or without slow waves. The increase of entropy waves is often considered as a possible trigger of coronal rain \cite{2020PPCF...62a4016A, 2022FrASS...920116A}. More details about the mixed properties of entropy and slow waves can be found in \cite{2021SoPh..296...96Z}. The set of discussed effects is known as a thermal misbalance effect, directly linked to the physics of thermal instabilities traditionally considered in heliospheric and interstellar plasma communities \cite{1965ApJ...142..531F}. The recent review on the thermal misbalance in application to the coronal plasma can be found in  \cite{2021PPCF...63l4008K}. Special attention deserves a modern look at the frequency-dependent damping of slow waves through the prism of the thermal misbalance \citet{2022MNRAS.514L..51K}.  It is shown that the thermodynamic activity modifies the relationship between the damping time and oscillation period, and it becomes a non-power-law function.

The thermal conduction also leads to the dispersion of the slow-wave phase velocity. In this case, the phase speed tends to the isothermal sound speed in the short period limit. The damping associated with thermal conduction also has a dependence on the wave period, namely, the higher harmonics decay faster. Furthermore, it also introduces the phase shift between the perturbations of plasma parameters. In fact, damping time and phase shifts can be expressed using the thermal conductivity coefficient. Usually, this coefficient is taken equal to the value proposed by Spitzer  \citep{1965pfig.book.....S}.  However, a series of studies by \cite{2002ApJ...580L..85O, 2015ApJ...811L..13W, 2019FrASS...6...57S,  2022ApJ...926...64O}  dedicated to the seismological determination of plasma transport coefficients propose the suppression of this coefficient. These seismological estimations are based on the assumption that the observed oscillation can be associated with the dynamics of the slow wave and application of the relationship between the thermal conduction coefficient and wave parameters.
In order to analyze the thermal conduction influence on the properties of slow waves and obtain required relationships, one can apply some simplifying approximations. In particular, the assumptions of weak \citep{2009A&A...494..339O, 2011ApJ...727L..32V, 2015ApJ...811L..13W, 2018ApJ...868..149K, 2022SoPh..297....5P} or strong \citep{2014ApJ...789..118K, 2020arXiv201110437D}  impact of thermal conduction are frequently used for this purpose. However, the results presented in \cite{2022FrASS...973664K}  show that by assuming that the thermal conductivity satisfies the value proposed by Spitzer, the dynamics of the fundamental slow-wave harmonic can be described with neither strong nor weak thermal conductivity limits (see Figure 2 in \cite{2022FrASS...973664K} for details). In this case, the general description without limitation on the thermal conduction coefficient is required. The latter can be obtained using, for example, the numerical solution of model equations.  

The current research aims to contribute to the problem of coronal seismology by non-adiabatic slow and entropy waves in coronal loops. In particular, we aim to derive explicit analytical conditions under which observed compression perturbations should be  associated with the  dynamics of slow waves, entropy waves, or their linear superposition. For this reason, we will obtain not a particular (for slow or entropy mode separately), but a full exact solution (for a combination of slow and entropy modes) to the evolutionary equation.  Furthermore, using the obtained solution, we will introduce some relationships between the thermal conduction coefficient and wave parameters, which can be applied to coronal seismology needs. Both these issues will be conducted without setting any restriction on the thermal conduction coefficient. 

Our manuscript is organized in the following way. In  Section  \ref{s:Model}, we introduce the analyzed MHD model, discuss the made assumptions, and indicate the main control parameter. Further, in Section \ref{s: Fourier Harmonics }, we briefly demonstrate the methodology used to derive the exact solution and show the difference between the cases of only thermal misbalance and only thermal conduction effect on wave dynamics. The description of the obtained exact solution for perturbations of various plasma parameters can be found in Section \ref{s: Exact sol}. We apply our theoretical results using the solar corona parameters in Section \ref{s: application}.
Finally, we summarize our results and introduce the main conclusions in Section \ref{s:Discussion}.

\section{Model and basic equations}
\label{s:Model} 

Further, we will analyze the linear evolution of slow MA and entropy waves. We assume that waves propagate  along magnetic field lines and apply the infinite magnetic field approximation. The compression viscosity is neglected. Thus, the evolutionary equation has the form shown below \citep{Zavershinskii2019}:
\begin{equation}\label{eq_general}
	\frac{\partial^3 a_1}{\partial t^3} -  c_{\mathrm{S}}^2 \frac{\partial^3 a_1}{\partial t \partial z^2}
	= \frac{\kappa}{\rho_0 C_V}\left(\frac{\partial^4 a_1}{\partial z^2\partial t^2} - c_{\mathrm{Si}}^2 \frac{\partial^4 a_1}{\partial z^4} \right) 
	-\frac{1}{\tau_\mathrm{V} }\left(\frac{\partial^2 a_1}{\partial t^2} - c_{\mathrm{S}Q}^2  \frac{\partial^2 a_1}{\partial z^2}\right).
\end{equation}
Here, $a$ means any variable (density $\rho$, temperature $T$, pressure $P$, or velocity $u$) describing the state of plasma. Hereafter, index \lq\lq 0\rq\rq\ means the stationary value of the parameter, and index \lq\lq 1\rq\rq\ indicates that the quantity is of the first order of smallness (i.e. $\rho_1/\rho_0 \sim T_1/T_0 \sim P_1/P_0 \sim u_1/c_\mathrm{S} \sim \epsilon \ll 1 $). The first term on the right-hand side (RHS) describes the influence of the field-aligned thermal conduction with the coefficient $\kappa$.  In turn, the second term on the RHS describes the effect of thermal misbalance on the wave dynamics. The characteristic timescales of the misbalance can be written as follows:
\begin{equation}
\tau_\mathrm{V}= \frac{C_\mathrm{V}}{\left( \partial Q  / \partial T \right)_{\rho}}, \quad
\tau_\mathrm{P}=\frac{C_\mathrm{P} }{\left( \partial Q  / \partial T \right)_{\rho}   - (\rho_0/T_0)  \left( \partial Q  / \partial \rho \right)_{T}},
\end{equation}
where, $C_V$ and  $C_P$ are specific heat capacities at constant volume and  pressure, respectively. The timescales $\tau_\mathrm{V}$, $ \tau_\mathrm{P}$ (  $\tau_2$, $ \tau_1$ in terms used in \cite{Zavershinskii2019}) are written using the derivatives of heat [$H$] and loss [$L$] function:
\begin{equation}\label{QFunc}
Q(\rho, T)=L (\rho , T ) - H(\rho , T ).
\end{equation} 
In equation (\ref{eq_general}), we also use characteristic speed:
\begin{equation}
c_{\mathrm{S}}= \sqrt{\frac{\gamma \mathrm{k}_\mathrm{B}  T_0}{m}}, \quad
c_{\mathrm{S}i}= \sqrt{\frac{\mathrm{k}_\mathrm{B}  T_0}{m}}, \quad
c_{\mathrm{S}Q}= \sqrt{\frac{\gamma_Q \mathrm{k}_\mathrm{B}  T_0}{m}}, \label{eq_speeds}
\end{equation}
where $\mathrm{k}_\mathrm{B}$ is the Boltzmann constant,  $m$ is the mean particle mass. The standard adiabatic (with the adiabatic index $\gamma=5/3$) and isothermal sound speeds are $c_{\mathrm{S}}$ and $c_{\mathrm{S}i}$, respectively. The speed $c_{\mathrm{S}Q}$  is the sound speed of wave propagation in the regime of strong thermal misbalance, prescribed by effective polytropic index $\gamma_Q= {\gamma\tau_\mathrm{V}}/{\tau_\mathrm{P}}$ \citep{1974A&A....37...65H, Molevich88}.

In the general case, when the characteristic timescales of the thermal misbalance and thermal conduction are of the same order, one has to solve equation (\ref{eq_general}) to give an exact description of the dynamics of some arbitrary linear perturbation. 
In \cite{2021SoPh..296...96Z}, we  considered the case of strong thermal misbalance and weak thermal conduction, i.e. neglected the first term on the RHS of equation (\ref{eq_general}) by assuming that $\tau_\mathrm{V}$ and $\tau_\mathrm{P}$ are much shorter than the characteristic time of thermal conduction. However, previous estimations of $\tau_\mathrm{V}$ and $\tau_\mathrm{P}$ showed that they vary in the fairly broad range, from a few minutes to a few hundred minutes, for typical combinations of coronal plasma parameters \citep[see e.g.][Figure 2]{2020A&A...644A..33K}. Hence, in the current study, the opposite problem is addressed. Namely, we consider the case of weak thermal misbalance and strong impact of thermal conduction. In other words, we will assume that the second term on the RHS of equation (\ref{eq_general}) can be neglected. 

Thus, governing evolutionary equation for slow MA and entropy waves in the plasma with strong impact of thermal conduction (strong in relation to thermal misbalance) can be written in the following dimensionless form: 
\begin{equation}\label{eq_lin_evol}
	\frac{\partial^3   \widetilde{a}_j}{\partial \tilde{t}^3} -  \gamma \frac{\partial^3 \widetilde{a}_j}{\partial \tilde{t} \partial \tilde{z}^2} 
	= \widetilde{d} \left( \frac{\partial^4 \widetilde{a}_j}{\partial \tilde{t}^2 \partial \tilde{z}^2} - \frac{\partial^4 \widetilde{a}_j}{ \partial \tilde{z}^4} \right).
\end{equation}

Here, we have introduced the dimensionless perturbation of plasma parameter $\widetilde{a}_j $. The index $j$ defines the parameter under study,  {i.e., it is used for symbols $\rho, P, T,$ and $u$}. In other words, we use the following values [$ \widetilde{a}_\rho= \rho_{1}/  \rho_0$] for density perturbation,   [$ \widetilde{a}_P = P_{1}/  P_0$] for pressure perturbation, [$ \widetilde{a}_T= T_{1}/ T_0$] for temperature perturbation, and  [$ \widetilde{a}_u= u_{1}/  c_{\mathrm{S}i} $] for velocity perturbation. We also use dimensionless coordinate [$ \tilde{z}= z/L$], and time [$ \tilde{t}= t/t_L, t_L = c_{\mathrm{S}i}/{L}$]. Here,   $L$   is the characteristic spatial scale.

Special attention should be paid to the dimensionless parameter $\widetilde{d}$:
\begin{equation}\label{eq_therm_cond_timescale}
\widetilde{d}= \frac{1}{\widetilde{\tau}_\mathrm{cond}} =  \frac{t_L}{\tau_\mathrm{cond}}, \quad \tau_\mathrm{cond}=  \frac{L^2 C_\mathrm{V} \rho_0}{\kappa},
\end{equation}
which is the reciprocal dimensionless characteristic timescale ${\widetilde{\tau}_\mathrm{cond}}$  attributed to the thermal conduction. It can be seen that $\widetilde{d}$ is the main control parameter that determines the spatiotemporal evolution of the perturbation. For clarity, we should mention that the timescale $\tau_\mathrm{cond}$ has a similar form as in \cite {2003A&A...408..755D}. In turn, the introduced  ${\widetilde{d}}$  is greater than value $d$ in \cite {2003A&A...408..755D} by $\gamma^{3/2}$ times.  

Further, we will obtain the exact solution of equation (\ref{eq_lin_evol}). In what follows, we will work in dimensionless units, so \textbf{the tilde sign above the quantities will be omitted.  The only exception will be made for $\widetilde{d}$, in order to distinguish it with respect to the differentiation operation.}

\section{Behaviour of the Individual Fourier Harmonics} \label{s: Fourier Harmonics }

In order to obtain the exact analytical solution for equation (\ref{eq_lin_evol}), we will follow the approach we used before in \citep{2021SoPh..296...96Z}. According to this approach, we will search for the solution for equation (\ref{eq_lin_evol}), using the Fourier method. In other words, we will use the substitution $ a\left(z,t \right) = \varphi\left(z\right) \psi\left(t\right)  $. It allows us to split equation (\ref{eq_lin_evol}) into two equations. The first one is the equation describing the dependence of the full solution on the coordinate, $\varphi\left(z\right)$:
\begin{equation}\label{spatial_eq}
	\frac{d^2 \varphi}{ d z^2} + k^2  \varphi = 0,
\end{equation}
This equation has to be solved using appropriate boundary conditions. In the current research, we will use reflecting  conditions which, in general, allows for the formation of standing slow waves. It means that one should apply  Dirichlet boundary conditions ($\varphi \left(0 \right) =  \varphi \left(l \right)   =0 $) for velocity equation and Neumann boundary conditions  ($d \varphi \left(0 \right)  /d  z = d \varphi \left(l \right)  / d z =0 $) for density, pressure and temperature equations.  In both cases, we obtain the set of eigenvalues $k$ defined by the harmonic number $ n $  as:
\begin{equation}\label{eigenvalue_eq}
	k = \frac{\pi n}{l}, n = 1,2,3,... 
\end{equation}
Here, $l$ is the length of the medium normalised to the characteristic length scale $L$. For definiteness, let us assume that $L$  is the loop length. 

In turn, the equation describing the dependence of the full solution on time $\psi\left(t\right)$ can be written in the form shown below:
\begin{equation}\label{temporal_eq}
	\frac{d^3 \psi}{ d t^3} + k^2(n)  \widetilde{d} ~ \frac{d^2 \psi}{ d t^2} +  k^2(n) \gamma   \frac{d \psi}{ d t} + k^4(n)  \widetilde{d}  \psi = 0.
\end{equation}
The solution to equation (\ref{temporal_eq})  can be found using the corresponding cubic equation. This cubic equation can be obtained by writing $d/d\,t \to i\omega$:
\begin{equation}\label{cubic_eq}
	\omega^3 - \mathrm{i} k^2(n) \widetilde{d} \omega^2 -  k^2(n) \gamma   \omega + \mathrm{i} k^4(n) \widetilde{d}  = 0,
\end{equation}
The cubic equation with complex coefficients (\ref{cubic_eq}) coincides with the general dispersion relation for slow and entropy modes in the plasma affected by thermal conduction only  (see e.g. \cite{2003A&A...408..755D}).  Similar to the case of the thermal misbalance influence only \citep{2021SoPh..296...96Z}, one can consider the roots of equation  (\ref{cubic_eq}) as complex frequencies $\omega_{1,2,3}$  of entropy and slow MA harmonics, which corresponds to real wavenumbers $k$. 

To describe the roots, let us introduce the discriminant $\Delta$ of Equation (\ref{cubic_eq}):
\begin{equation}\label{discriminant_eq}
\Delta=-108(R^3+U^2), 
\end{equation}
where  $R$ and $U$ are real coefficients defined by wavenumbers and characteristic dimensionless parameters of plasma:
\begin{equation}\label{notation_discr_eq}
R =k^2 \frac{3  \gamma - k^2 \widetilde{d}^2 }{9},~~
U = k^4 \widetilde{d} ~ \frac{ 2 k^2  \widetilde{d}^2 - 9 \left( \gamma - 3  \right)}{ 54}.\nonumber
\end{equation}
Thus, the roots of the dispersion relation (\ref{cubic_eq}) according to  Cardano's formula take the form shown below:

\begin{align}\label{roots_eq} 
	& \omega_1  = \mathrm{i} \left(-\frac{k^2\widetilde{d}  }{3}+A+B\right),\nonumber \\ 
	&\omega_2 =   \frac{A-B}{2} \sqrt{3}  -\mathrm{i}\left(\frac{k^2\widetilde{d} }{3}+\frac{A+B}{2}\right), \\	
     &\omega_3 =  - \frac{A-B}{2} \sqrt{3} -\mathrm{i}\left(\frac{k^2\widetilde{d} }{3}+\frac{A+B}{2}\right),  \nonumber  
\end{align}
where 
\begin{equation}\label{notation_roots_eq}
	A = \sqrt[3]{-U + \sqrt{-\Delta/108}} ,~~B = - R / A.\nonumber
\end{equation} 

Analyzing equation (\ref{discriminant_eq}) one can show that for any  $k$ and $\widetilde{d} $ the  discriminant is negatively defined $\Delta < 0$. This means that regardless of plasma parameters, the dispersion relation (\ref{cubic_eq}) always has one purely imaginary root and two complex conjugate roots.

The purely imaginary root can be associated with the decrement of the entropy mode harmonics:
\begin{equation}\label{im_frq_ent_eq}
	\omega_\mathrm{EI} = -\frac{k^2\widetilde{d} }{3}+A+B. 
\end{equation}
In turn, the complex conjugate roots $\omega_{2,3}=\pm\omega_\mathrm{AR}-\mathrm{i}\omega_\mathrm{AI}$ can be associated with two oppositely propagating decaying slow waves. The imaginary part $\omega_\mathrm{AI}$ (i.e. decrement of slow waves) and real part $\omega_\mathrm{AR}$ are as follows:
\begin{equation}\label{re_im_frq_ac_eq}
	\omega_\mathrm{AI} =\frac{k^2 \widetilde{d} }{3} + \frac{A+B}{2},~~\omega_\mathrm{AR}=\frac{A-B}{2} \sqrt{3}.
\end{equation}
This is significantly different from the case of misbalance only considered by \citep{2021SoPh..296...96Z}, in which the properties of slow and entropy waves can be mixed. In other words, when some harmonics of slow waves can become non-propagating and evolve in a similar way to entropy mode harmonics. In the plasma with the strong impact of thermal conduction, there is no such possibility. The entropy modes are always decaying and non-propagating and two slow waves are always propagating and decaying.

\section{Exact Solution} \label{s: Exact sol}

As we have mentioned before, the solution for equation (\ref{eq_lin_evol}) can be written as the sum of harmonics of entropy and slow modes. Let us describe the solution for the case of Neumann reflecting boundary conditions (i.e. for density, pressure, or temperature perturbations). For definiteness, we will consider density perturbation. Then, the solution for the $n$th harmonic  can be written as follows:
\begin{eqnarray}\label{rhon_delt_min_eq}
		&a_{\rho n}\left(z,t\right) = C_{1 \rho  n} \mathrm{e}^{\omega_\mathrm{EI}t} \cos \left( k z\right) + \\
		&C_{0\rho n} \mathrm{e}^{\omega_\mathrm{AI}t}   \left[\cos \left(\omega_\mathrm{AR} t + k z  - \phi_ {\rho n} \right)+ \cos \left(\omega_\mathrm{AR} t - k z  - \phi_ {\rho n} \right) \right],  \nonumber
\end{eqnarray}
	where 	
\begin{equation}\label{not_rhon_delt_min}
		C_{0 \rho  n}=\frac{\sqrt{C_{2 \rho n}^2+C_{3 \rho  n}^2}}{2},~~\phi_ {\rho n} = \arctan \left(\frac{C_{3 \rho n}}{C_{2 \rho n}}\right).
\end{equation}
The first and second terms in equation (\ref{rhon_delt_min_eq}{}) correspond to non-propagating entropy-mode harmonic and two slow-mode harmonics propagating in opposite directions. To define the magnitudes of entropy-mode $C_{1 \rho  n}$ and slow-mode $C_{0 \rho  n}$ harmonics, one has to solve the following set of linear equations,
\begin{equation}\label{const_mat_delt_min}
		\begin{pmatrix} 1 & 1 & 0 \\ \omega_\mathrm{EI} & -\omega_\mathrm{AI} & \omega_\mathrm{AR} \\ \omega_\mathrm{EI}^2 & \left(\omega_\mathrm{AI}^2-\omega_\mathrm{AR}^2\right) & -2\omega_\mathrm{AR}\omega_\mathrm{AI} \\  \end{pmatrix}   		\begin{pmatrix} C_{1  \rho  n}  \\ C_{2  \rho n} \\ C_{3  \rho n}  \end{pmatrix}  	 =		\begin{pmatrix} I_{1n}   \\  I_{2n} \\ I_{3n}  \end{pmatrix} .
\end{equation}
	The integrals in the right-hand side $I_{1n}$, $I_{2n}$, and $I_{3n}$  are prescribed by the initial perturbation $\rho_\mathrm{in}(z,0)$ and the derivatives $\at{(\partial \rho(z,t) /  \partial t)}{t=0}$, and $ \at{(\partial^2 \rho(z,t) /  \partial t^2)}{t=0}$ as
 	\begin{align}
		&I_{1n}  = \frac{2}{l} \int_0^l \rho_\mathrm{in}(z,0) \cos \left( k z\right) \mathrm{d}z,
        \qquad\qquad I_{2n}  = \frac{2}{l} \int_0^l \at{\frac{\partial \rho(z,t)}{\partial t}}{t=0}  \cos \left( k z\right) \mathrm{d}z, \label{init_integral} \\
		&\qquad\qquad\qquad I_{3n} = \frac{2}{l} \int_0^l \at{\frac{\partial^2 \rho(z,t)}{\partial t^2}}{t=0} \cos \left( k z\right) \mathrm{d}z. \nonumber
	\end{align}
The non-oscillating and non-propagating background is defined by the expression shown below:
	\begin{equation}\label{rho0_delt_0_eq}
		a_{\rho 0}\left(z,t\right) =  I_{10}  = \frac{1}{l} \int_0^l \rho_\mathrm{in}(z,0) \mathrm{d}z.
	\end{equation}
To construct the exact solution of equation~(\ref{eq_lin_evol}), one should apply the superposition principle. This means, that the full compressive perturbation can be described by the sum of non-oscillating and non-propagating background (\ref{rho0_delt_0_eq}) and all harmonics from $n=1$ to infinity, using equation (\ref{rhon_delt_min_eq}):
	\begin{equation}\label{eq_solution}
	a_{\rho} \left(z,t\right)  = 	a_{\rho 0}\left(z,t\right) + \sum\limits_{n=1}^{\infty} a_{\rho n}\left(z,t\right).
	\end{equation}
The solution for velocity perturbation, which requires Dirichlet boundary condition, can be constructed in a similar way. There are two main differences in solutions. First, the 
cosine function in equations(\ref{rhon_delt_min_eq})  and (\ref{init_integral})  must be replaced by sine. Second, the background is zero as we assume that the plasma under consideration is non-propagating.

\section{Applications of the exact solution} \label{s: application}

The obtained exact solution (\ref{eq_solution}) has a number of applications for analyzing the properties and evolution of compression perturbations in plasma.  As we have mentioned previously, it allows us not only to follow the dynamics of the full solution but also to monitor the evolution of the eigenmodes that determine it. Moreover, we also can analyze how any harmonics of slow and entropy modes evolve. Further, in this section, we will apply the obtained exact solution to illustrate some evolution ways of the localised initial perturbation and discuss some results revealed by means of it.

\subsection{Spatio-temporal evolution of entropy and slow MA modes}

 First, in order to analyze the spatio-temporal evolution, we have to define the initial perturbation. In this work, we will follow the way we applied in  \citep{2021SoPh..296...96Z}, to enable the reader to compare results. In other words, we  will consider some  Gaussian pulse, which perturbs plasma density, pressure, and temperature:
\begin{align}\label{init_gaussian} 
	&a_{\rho,in}\left(z,0\right)= A_{\rho} \exp\left[-\left(z-z_0\right)^2 / w\right],\quad a_{P,in}\left(z,0\right)= A_{P} \exp\left[-\left(z-z_0\right)^2 / w\right],\nonumber \\ 
	&a_{T,in}\left(z,0\right) = a_{P,in}\left(z,0\right) - a_{P,in}\left(z,0\right), \quad 	a_{u,in}\left(z,0\right) = 0.  
\end{align}
Here,  $A_{\rho}$ and $A_P$ are dimensionless magnitudes of the density and pressure variations; $w$ and $z_0$ are the effective width and position of the perturbing pulse, respectively.

\begin{table}[h!]
\begin{center}
\begin{tabular}{|c|c|c|c|c|}
\hline
Type                                & Length                & Temperature           & Density               & ${\widetilde{\tau}_\mathrm{cond}}$                     \\
                                    & $[10^9 \mathrm{cm}]$             & [MK]                  & $[10^9 \mathrm{cm^{-3}}]$        &                        \\ \hline
Bright Points                       &  $0.1-1$              & $2$                   &  $5$                  &  $\sim   0.3-3$        \\ \hline
Active Region                       &  $1-10$               & $3$                   &  $1-10$               &     $\sim   0.3-27.6$   \\ \hline
Giant Arches                        &   $10-100$            &  $1-2$                &    $0.1-11$            &   $\sim   2.6-26.4$    \\ \hline
\multicolumn{1}{|l|}{Flaring loops} & \multicolumn{1}{l|}{$~~1-10$  } & \multicolumn{1}{l|}{$~~~~~>10$} & \multicolumn{1}{l|}{$~~~~~>50$} & \multicolumn{1}{l|}{$~~~~~\gtrsim 1.2$} \\ \hline
\end{tabular}
\caption{ Estimations of the characteristic timescale ${\widetilde{\tau}_\mathrm{cond}}$ (\ref{eq_therm_cond_timescale}) calculated for typical coronal loop parameters taken from \cite{2014LRSP...11....4R}.
	} \label{tab_param}
 \end{center}
\end{table}

Secondly, the ranges of control parameters should be specified. The main control parameter in this work is the characteristic timescale attributed to the thermal conduction $\tau_\mathrm{cond}$  or its reciprocal  $\widetilde{d}$. To estimate the range of $\tau_\mathrm{cond}$, we use typical coronal loop parameters from Table 1 of \cite{2014LRSP...11....4R}. Our estimations of the characteristic timescale $\tau_\mathrm{cond}$ calculated for various coronal loop parameters are shown in Table \ref{tab_param}.
One may notice that the timescale $\tau_\mathrm{cond}$  also vary in the  fairly broad range.  The dimensional characteristic time (\ref{eq_therm_cond_timescale}), can be an order of magnitude or several orders of magnitude greater or less than the characteristic travel time of isothermal sound, even for loops belonging to the same type. The latter applies to loops in active region and bright points.  Under coronal temperatures, the increase in characteristic thermal conduction  time  is primarily determined by an increase in loop length and density.

\setcounter{figure}{1}
\setcounter{subfigure}{0}
\begin{subfigure}
\setcounter{figure}{1}
\setcounter{subfigure}{0}
    \centering
    \begin{minipage}[b]{\textwidth}
        \includegraphics[width=\linewidth]{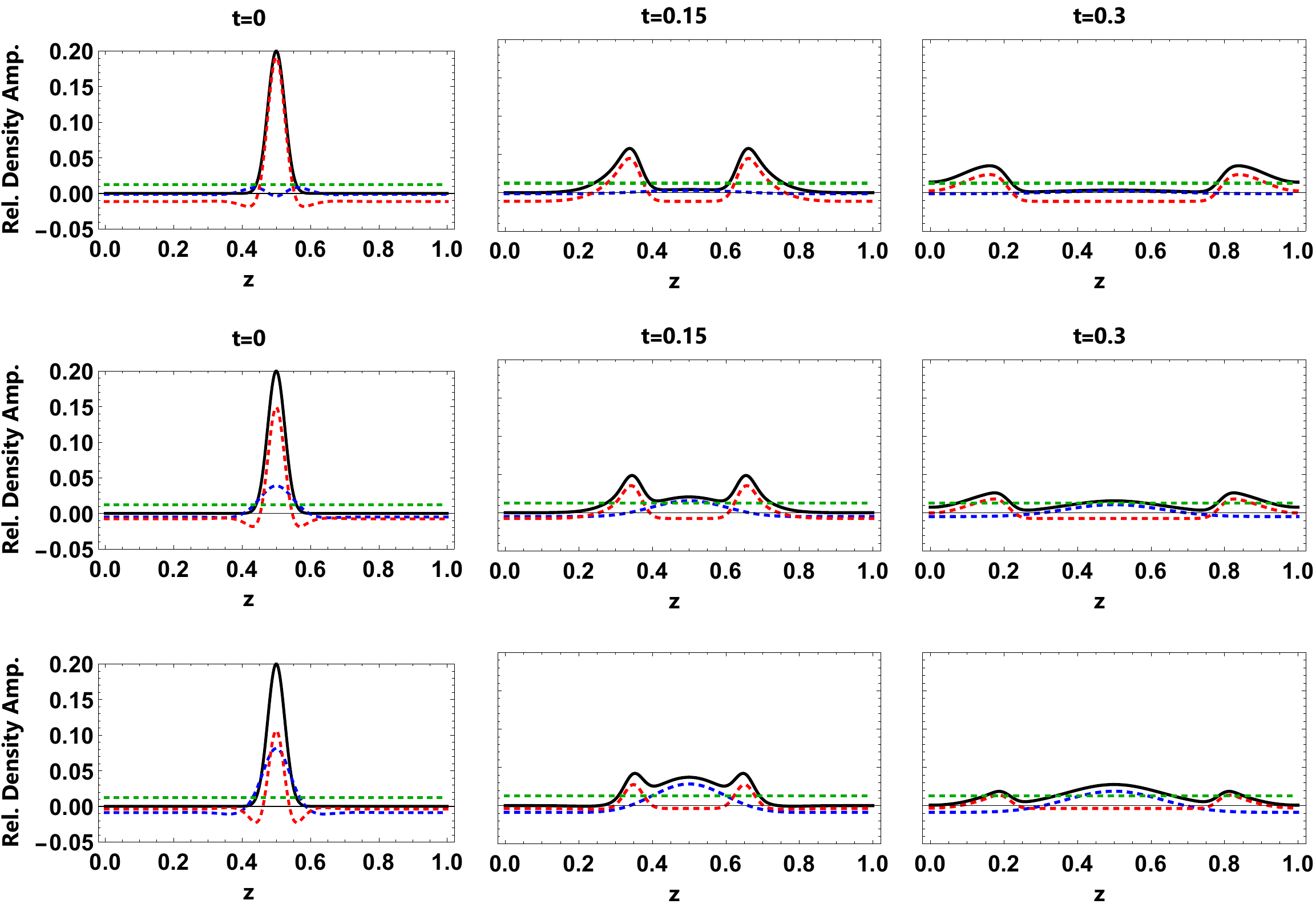}
        \caption{Evolution of the initial Gaussian perturbation with width $w=l/40$ situated at $z_0 = l/2$ . The characteristic time is set to $\tau_\mathrm{cond} = 25$. The \textit{top}, \textit{middle} and \textit{bottom} rows correspond to the ratio of magnitudes $A_{P}/A_{\rho} $ equals $1.5, 1$ and $0.5$, respectively. \textit{Left}, \textit{middle}, and \textit{right} columns indicate the solutions at $t = 0, t = 0.15,$ and $t = 0.3$ of the computational time, respectively.  The \textit{ black } solid line corresponds to the full solution (sum of solutions for one entropy and two slow MA modes). The \textit{red} and \textit{blue }dashed lines indicate the sum of two slow MA modes and one entropy mode, respectively. The \textit{green } dashed line corresponds to the non-oscillating and non-propagating background. }
        \label{fig:Subfigure 1}
    \end{minipage}  
   
\setcounter{figure}{1}
\setcounter{subfigure}{1}
    \begin{minipage}[b]{\textwidth}
        \includegraphics[width=\linewidth]{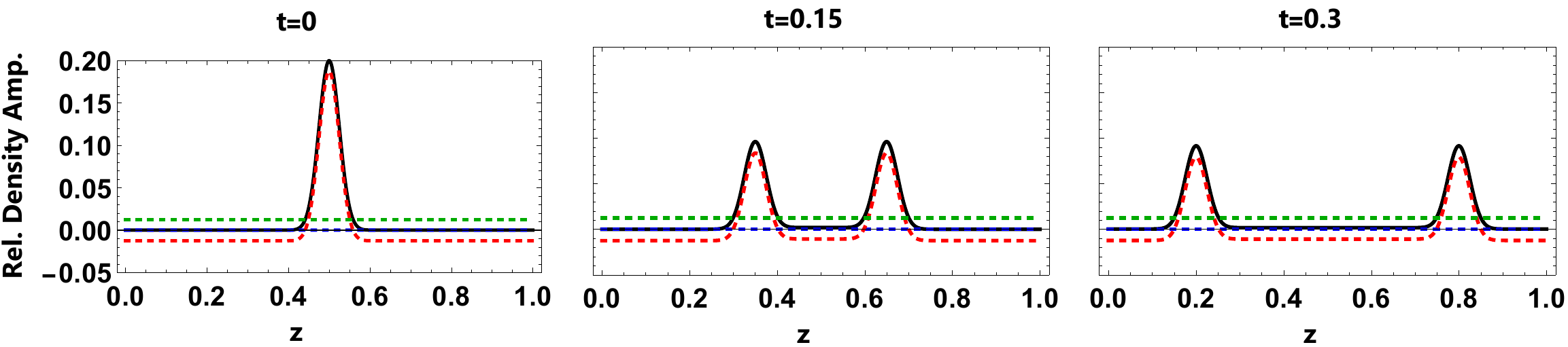}
        \caption{Evolution of the initial Gaussian perturbation  with width $w=l/40$ situated at $z_0 = l/2$. The characteristic time is set to $\tau_\mathrm{cond} = 1$. Calculations are made for the ratio of magnitudes $A_{P}/A_{\rho} $ equal to $ 1$. }
        \label{fig:Subfigure 2}
    \end{minipage}

\end{subfigure}

As we have mentioned above, in this manuscript, we would like to draw attention to the fact, that in the solar corona, the initiated compression perturbation should not always be associated with the slow wave only.  In the general case, the initiated perturbation is polytropic. In other words, the ratio of pressure and density magnitudes $A_{P}/A_{\rho} $ is not necessarily equal to $\gamma=5/3$ like in the slow mode perturbation and depends on the initiating mechanism. It is quite intuitive that the distribution of energy between the entropy mode and slow modes should be defined by the form and type of initial condition. To illustrate this fact, we use the initial perturbation (\ref{init_gaussian}) and consider three different values $1.5, 1$ and $0.5$  of ratio $A_{P}/A_{\rho} $. It can be seen in Figure \ref{fig:Subfigure 1} that an increase in the difference with respect to $\gamma=5/3$ (i.e., decrease in the perturbation adiabaticity) increases the contribution of the entropy mode to the full solution (blue lines). One may also notice the asymmetry of slow mode perturbation growing in time, this is a result of slow wave dispersion caused by thermal conduction. Longer harmonics propagate faster towards the long wavelength limit $c_{\mathrm{S}}$ (adiabatic sound speed)  than shorter harmonics tending to $c_{\mathrm{Si}}$ (isothermal sound speed) (\ref{eq_speeds}).

However, our analysis also revealed a very interesting result, namely, that the distribution of energy between modes depends on the value of thermal conduction and the coronal loop parameters (i.e. on $\tau_\mathrm{cond}$). We show the temporal evolution of Gaussian pulse in the plasma with low $\tau_\mathrm{cond}$ in Figure \ref{fig:Subfigure 2}. For illustration,  we choose the isothermal initial condition ($A_{P}/A_{\rho} = 1$). The calculations revealed that in the plasma with low $\tau_\mathrm{cond}$, the initiating mechanism becomes insignificant and the full solution is defined primarily by the slow mode. 

In order to verify the obtained exact solution (\ref{eq_solution}), we compare it with the numerical solution of equation (\ref{eq_lin_evol}). The animation, which shows the construction of an exact solution by summation of harmonics can be found in the Supplementary material. In practice, accounting for the first $\sim 50$ harmonics in equation (\ref{eq_solution}) allows for the reproduction of the considered full solution with sufficient accuracy.

\subsection{Partition of energy between entropy and slow MA modes}
\begin{figure}
	\begin{center}
		\includegraphics[width=0.7\linewidth]{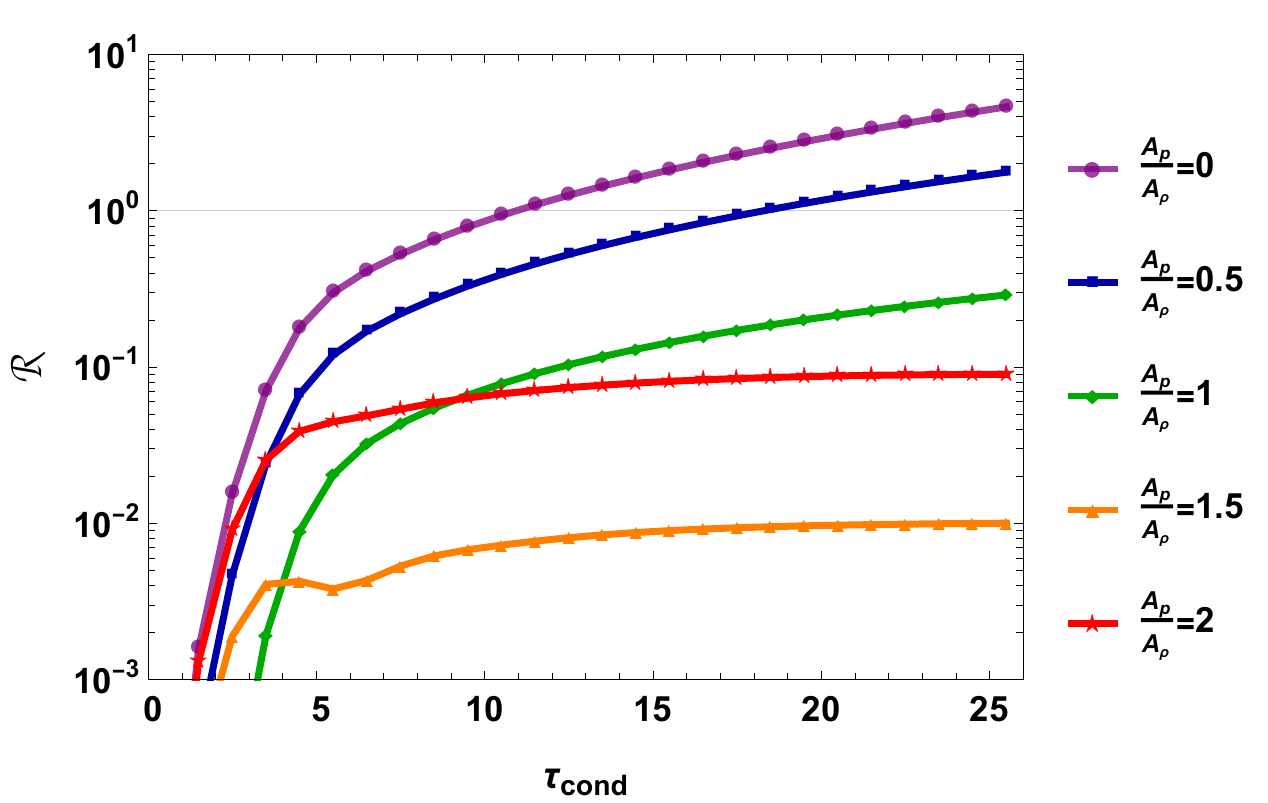}
	\end{center}
	\caption{Dependence of the ratio ${\mathcal R}$  {(\ref{AmpRatio_Eq})} on dimensionless  characteristic timescale $\tau_\mathrm{cond}$  {(\ref{eq_therm_cond_timescale})}  calculated for different values  magnitude  ratio $A_{P}/A_{\rho}$.
	}
	\label{SolutFig}
\end{figure}
In this Section, we will discuss the issues, when taking into consideration the influence of the entropy mode (even at the stage of perturbation initialization) becomes highly important. In order to answer this question, we aim to calculate the partition of the total (i.e. integrated over all harmonics) initial energy between the modes. For this purpose, we apply the formulation used before in  \citep{2021SoPh..296...96Z} and estimate the  ratio of the total initial energies gained by the entropy and slow MA modes, respectively, from the initial Gaussian pulse as  
\begin{equation}\label{AmpRatio_Eq}
	{\mathcal R} = \sum\limits_{n=1}^{\infty} C_{1n}^2/ \sum\limits_{n=1}^{\infty} 4 C_{0n}^2 =\frac{Es}{As}.
\end{equation}

In our calculations, we consider how the ratio ${\mathcal R}$ depends on the characteristic timescale $\tau_\mathrm{cond}$ and the type of initial perturbation,  defined by ratio $A_{P}/A_{\rho}$. The results of our estimations are shown in Figure \ref{SolutFig}. One may notice the two main features, which have been proposed using the analysis of the spatio-temporal evolution of complete perturbation.

The first one concerns the relation between the initiation mechanism of the original signal and the energy distribution. It is seen from  Figure \ref{SolutFig}, that the slow mode tends to be the dominant part of the full solution,  when the initial perturbation tends to be adiabatic ($A_{P}/A_{\rho} \rightarrow \gamma=5/3$). The closest to this condition shown in Figure \ref{SolutFig} is the orange curve, corresponding to $A_{P}/A_{\rho} = 1.5$. It is seen, that for any $\tau_\mathrm{cond}$, the impact of entropy mode is negligible in this case, for illustration we refer to previously shown results in the top row in Figure \ref{fig:Subfigure 1}. The greater the difference of magnitude ratio $A_{P}/A_{\rho} $  from value $1.66$, the greater impact of entropy mode in the full solution. It can be seen using Figure \ref{SolutFig} by comparing the  behavior of orange  ($A_{P}/A_{\rho} = 1$.5) and green  and  red curves ($A_{P}/A_{\rho} = 1$ and  $A_{P}/A_{\rho} = 2$, respectively). For an illustration of this effect, we refer to Figure \ref{fig:Subfigure 1}, namely, the middle and bottom rows.

The second feature concerns the relation between the energy distribution and the characteristic timescale $\tau_\mathrm{cond}$. It is clearly seen that the entropy mode becomes more important for the greater value of $\tau_\mathrm{cond}$  and that the dependence of ${\mathcal R}$ on $\tau_\mathrm{cond}$ is highly non-linear. One can notice that for some initialization mechanisms, the entropy mode can become not only comparable to the slow mode but the dominant part of the full solution (see for example purple and blue curves in Figure \ref{SolutFig}).  For an illustration of the mentioned feature, we recommend to compare the results in Figure \ref{fig:Subfigure 1}  and \ref{fig:Subfigure 2}).  A similar sensitivity of the apparent entropy mode excitation to both the perturbation type and characteristic value of thermal conduction has been demonstrated numerically by \citet{2022FrASS...973664K}.

Further, using the obtained results and estimations of characteristic time $\tau_\mathrm{cond}$ (see Table \ref{tab_param}), we propose,  when taking into consideration the influence of the entropy mode becomes important.  In particular, an analysis of the compression perturbations in the bright points can be carried out under the assumption that the perturbation is a slow wave with sufficiently high accuracy. The same conclusion can be applied to the perturbation in rarefied and short coronal loops in the active region, for giant arches and flaring loops. In turn, in the longer and denser coronal loops, the influence of the entropy mode becomes more and more significant or even becomes the main feature of the perturbation evolution. And one of these specific cases, will be discussed in the following subsection.

\subsection{Phase shifts}
\begin{figure}
	\begin{center}
		\includegraphics[width=0.7\linewidth]{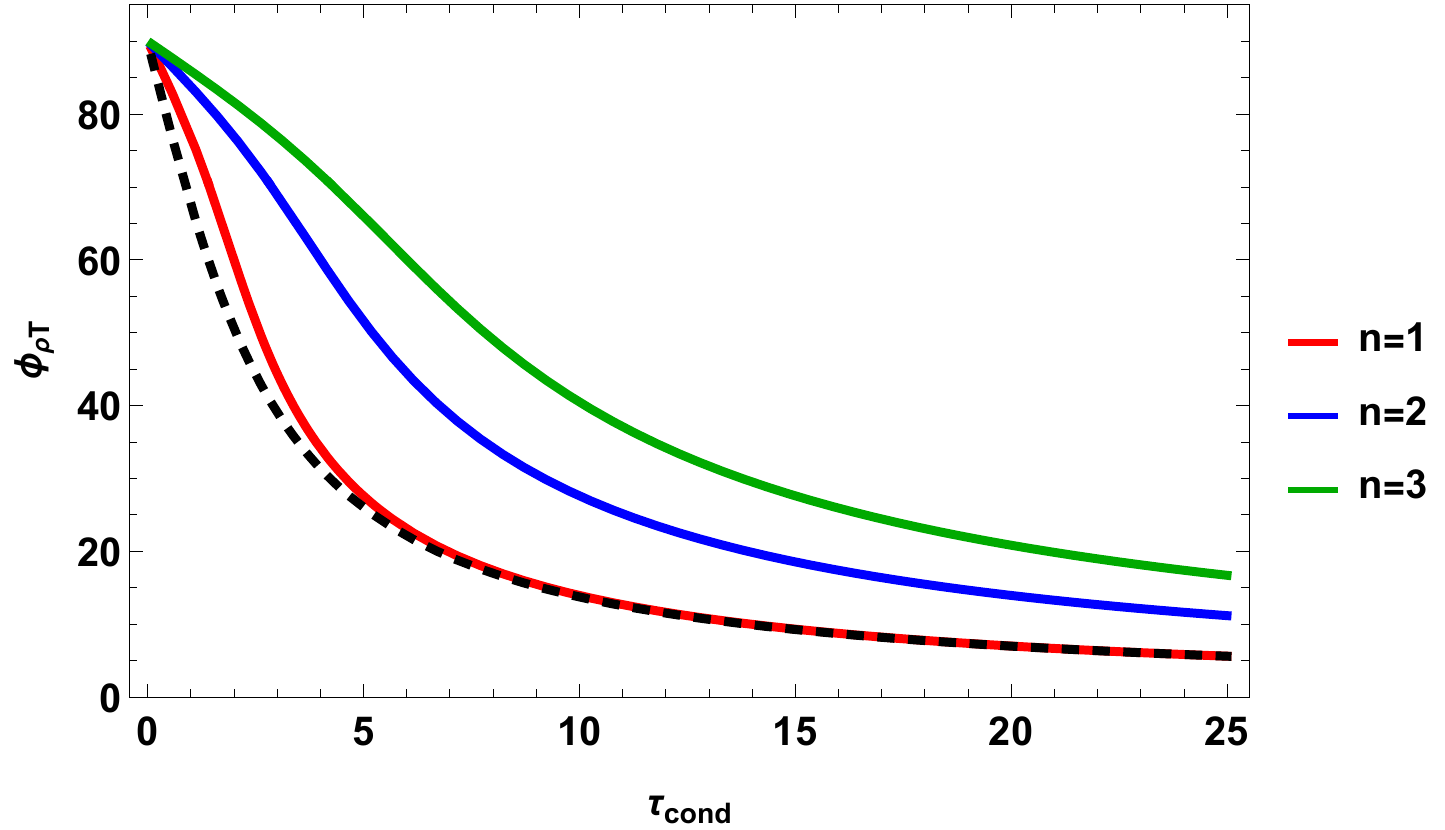}
	\end{center}
	\caption{The phase shifts $\phi_{\rho T} ($\ref{eq_phaseshift_T_Rho}) between density  and  temperature perturbations  {as a function of dimensionless  characteristic timescale $\tau_\mathrm{cond}$ (\ref{eq_therm_cond_timescale})}. The \textit{red}, \textit{blue}, and \textit{green} colors are for the first(fundamental), second, and third harmonic, respectively. The dashed line indicates the dependence described by the expression for the phase shift of fundamental harmonic obtained using assumptions of the weak impact of thermal conduction (see e.g.  \cite{2009A&A...494..339O}).
	}
	\label{f_phaseshifts}
\end{figure}

In this subsection, we introduce the expressions for phase shifts between perturbations of various plasma parameters. 

In order to derive the relationships for the phase shifts in the slow mode, we derive the full exact solution for all considered plasma parameters, namely, for density, temperature, pressuse and velocity perturbation. Using obtained solutions, we introduce the phase shift between temperature and density by calculating the following difference: 
\begin{equation}\label{eq_phaseshift_T_Rho}
\phi_{\rho T} =   \phi_{\rho n}  - \phi_{Tn} =	\arctan 
\left( 
\frac{\left( \omega_\mathrm{AI}^2  + \omega_\mathrm{AR}^2 \right) \sin{2 \phi_\mathrm{\rho u} }  }
{ \left( \omega_\mathrm{AI}^2  + \omega_\mathrm{AR}^2 \right) \cos{2 \phi_\mathrm{\rho u} }  + k^2  } 
\right).
\end{equation}
where $n$ is the harmonic number.  Here, we use the following notation
\begin{equation}\label{eq_phaseshift_Rho_u}
\phi_{\rho u} = \phi_{\rho n}  - \phi_{un}  =	\arctan \left( \frac{- \omega_\mathrm{AR} }{\omega_\mathrm{AI}} \right),
\end{equation}
which is the phase shift between density and velocity perturbation in the slow wave. In order to estimate the phase shifts, the real $\omega_\mathrm{AR}$ and imaginary $\omega_\mathrm{AI}$ parts of the slow wave frequencies have to be calculated. For this purpose, one can use previously introduced analytical expressions (\ref{re_im_frq_ac_eq}) following from dispersion relation (\ref{cubic_eq}) or the  numerical solution of the dispersion relation (\ref{cubic_eq}). 

It should be noted that obtained relationships (\ref{eq_phaseshift_T_Rho}) and (\ref{eq_phaseshift_Rho_u}) are valid not only for the fundamental harmonic but can be applied for the estimation of the phase shift of any arbitrary harmonic. Since these expressions follow directly from the exact solution, they also have no restrictions on the value of the thermal conduction coefficient. Thus, the expressions (\ref{eq_phaseshift_T_Rho}) and (\ref{eq_phaseshift_Rho_u}) are generalisations of the previously obtained relationship for the fundamental harmonic \citep[see e.g.][Eq. 51]{2021SSRv..217...34W}.

Let us analyze the dependence of the phase shift between temperature and density perturbations for the first three harmonics on characteristic timescale  $\tau_\mathrm{cond}$. Our calculations are shown in Figure \ref{f_phaseshifts}. One can notice, that regardless of the number of harmonics, the phase shift  {$\phi_{\rho T}$ (\ref{eq_phaseshift_T_Rho})} tends to $90^\circ$  as $\tau_\mathrm{cond} \rightarrow 0$ and tends to $0^\circ$ as $\tau_\mathrm{cond} \rightarrow \infty$. However, an increase in the harmonic number leads to an increase in the phase shift compared to the fundamental harmonic. Here, we also compare our results with the well-known expression for the phase shift of fundamental harmonic obtained using assumptions of the weak impact of thermal conduction (see e.g.  \citep{2009A&A...494..339O}), which is shown by the dashed curve in Figure \ref{f_phaseshifts} and can be written as $  \arctan \left( \pi /\tau_\mathrm{cond} \sqrt{\gamma} \right) $ using introduced dimensionless units. The approximate solution works well in the area of its applicability, however, with a strong effect of thermal conductivity, differences appear.




For example, \cite{2019MNRAS.483.5499K}  observed the event of quasi-periodic pulsations (QPP) which has been associated with the second harmonic of a standing slow MA wave in a flaring loop. The oscillations with period $P \approx 74–80~ \mathrm{s}$ have been detected in the loop with length $L \approx 17–22~  \mathrm{Mm}$,  temperature  $T \approx 1.1–1.6~  \mathrm{MK}$. The estimated  plasma density is $N  \approx 1.2–1.5 \times  10^{11}~  \mathrm{cm}^{-3}$ and assumed magnetic field is $B  \approx 100–150~  \mathrm{G}$. The phase shift observed between temperature and density perturbations has been estimated to equal $\phi_{\rho T}\approx 180^\circ$.  It follows, from the result shown in  Figure \ref{f_phaseshifts}, that such a large value of the phase shift cannot be described by the thermal conduction, even for the second harmonic of a slow wave. However,  for the given loop parameters the estimated dimensionless characteristic timescale $\tau_\mathrm{cond} \approx 220$ (\ref{eq_therm_cond_timescale}). According to the results shown in Figure \ref{SolutFig}, this value of $\tau_\mathrm{cond} $ implies the significant impact of entropy mode in the perturbation evolution. In this case, the observed compression perturbation cannot be unambiguously associated with the slow mode. In other words, the use of the expressions for slow mode phase shifts  (\ref{eq_phaseshift_T_Rho}) and (\ref{eq_phaseshift_Rho_u}) for the interpretation of observations may not be justified.

\section{Summary and conclusion}
\label{s:Discussion}

The obtained exact solution (\ref{eq_solution}) presents the effect of thermal conduction on the evolution of the compression perturbation in the non-adiabatic plasma of the hot solar corona. This solution gives us comprehensive information about the spatio-temporal dynamics of some initiated variations of plasma parameters. Moreover, it allows us not only to follow the dynamics of the full solution but also to analyze the eigenmodes that determine it, namely, entropy and slow mode. In particular, we can check how justified the association of the observed oscillations with the evolution of only slow waves is. Further, we summarise results, which have been revealed using the exact solution obtained.

\begin{itemize}

	\item  The theory presented is developed for any values of the characteristic  timescale $\tau_\mathrm{cond}$ associated with thermal conduction. Thus, the performed quantitative analysis extends the predictions made using assumptions of  weak \citep{2009A&A...494..339O, 2011ApJ...727L..32V, 2015ApJ...811L..13W, 2018ApJ...868..149K, 2022SoPh..297....5P} or  strong \citep{2014ApJ...789..118K, 2020arXiv201110437D}  impact of thermal conduction. It also generalises the phase-shift theory constructed earlier with no constraints on thermal conduction \citep[see e.g.][Eq. 51]{2021SSRv..217...34W} for the fundamental harmonic to the case of an arbitrary harmonic number.

	\item  Analyzing the behaviour of individual Fourier harmonic, we solve the general dispersion  relation (\ref{cubic_eq}) for slow and entropy modes  in the plasma affected by thermal conduction only (see e.g. \cite{2003A&A...408..755D}. Our analysis of its solution (\ref{roots_eq}) reveals that in contrast to the thermal misbalance \citep{2021SoPh..296...96Z}, the thermal conduction gives no possibility for properties of entropy mode and slow mode to be mixed. The slow waves are always propagating and decaying and entropy modes are always decaying  and non-propagating for any value of thermal conduction and harmonic number.
	\item In this manuscript, we analyze some possible evolution ways of Gaussian compression perturbation  (see Figs. \ref{fig:Subfigure 1} and \ref{fig:Subfigure 2}). It has been shown, that the energy fraction attributable to the slow wave in the full solution can vary not only due to a change in the initialization mechanism but also due to a change in the characteristic timescale $\tau_\mathrm{cond}$ attributed to the thermal conduction and loop parameters. Our estimations of the ratio of the total initial energies gained by the entropy and slow MA modes in an initial Gaussian pulse are shown in Figure \ref{SolutFig}. In order to propose for the oscillations in which coronal loops it is important to take into account the entropy mode on a par with the slow wave, we estimated the characteristic time   $\tau_\mathrm{cond}$ for typical coronal loop parameters \citep{2014LRSP...11....4R}. It follows that in the bright points, the compression perturbation is totally defined by the slow mode only. However, for the long and dense loops in the active region, flaring loops, and giant arches the entropy mode can become not only comparable but the dominant part of perturbation. 

	\item The obtained exact solution allows us to derive the expressions (\ref{eq_phaseshift_T_Rho}) and (\ref{eq_phaseshift_Rho_u}) for the phase shifts between perturbations of various plasma parameters, without limitation on the harmonic number and thermal conduction impact. The phase shift between temperature and density $\phi_{\rho T}$ (\ref{eq_phaseshift_T_Rho}) calculated for the first three harmonics and various values of characteristic timescale  $\tau_\mathrm{cond}$ is shown in Figure \ref{f_phaseshifts}. It is seen that for any values of thermal conduction and harmonic number, the phase shift is between  $90^\circ$  and $0^\circ$. Such a conclusion is in agreement with the results obtained using assumptions of the weak impact of thermal conduction for the fundamental harmonic only. (see e.g. \cite{2009A&A...494..339O}). Using our analytical results, we show that the approximate solution for $\phi_{\rho T}$ works well in the area of its applicability, however, with a strong effect of thermal conductivity, differences appear.  An increase in the harmonic number leads to an increase in the phase shift compared to the fundamental harmonic. Comparing our estimations of the second harmonic phase shift and observations presented by \cite{2019MNRAS.483.5499K}, we should admit that the observed value $\phi_{\rho T}\approx 180^\circ$ are out of range of values prescribed by thermal conduction effect.  However, our estimation of characteristic thermal conduction timescale  for the given loop parameters gives value $\tau_\mathrm{cond} \approx 220$. According to the results shown in Figure \ref{SolutFig}, this means that observed compression perturbation may not be unambiguously associated with the slow mode. Thus, the use of the expressions for slow mode phase shifts in their pure form (\ref{eq_phaseshift_T_Rho}), and (\ref{eq_phaseshift_Rho_u}) may not be justified.
 
\end{itemize}

As we have mentioned previously, the main  limitation of the conducted research is the neglecting of the thermal misbalance effect. In our future work, we plan to consider the combined effect of thermal conduction and thermal misbalance. In other words, we are aiming for derivation of exact solution of equation (\ref{eq_general}). The solution of such a general problem will improve the accuracy of the determination coronal heating function using compression perturbation. Nevertheless, in our opinion for the regions where the effect of thermal misbalance is weak, the presented solution (\ref{eq_solution}) and expression following from it  are a good help for coronal seismology problems.

\section*{Conflict of Interest Statement}

The authors declare that the research was conducted in the absence of any commercial or financial relationships that could be construed as a potential conflict of interest.

\section*{Author Contributions}

The Author Contributions section is mandatory for all articles, including articles by sole authors. If an appropriate statement is not provided on submission, a standard one will be inserted during the production process. The Author Contributions statement must describe the contributions of individual authors referred to by their initials and, in doing so, all authors agree to be accountable for the content of the work. Please see  \href{https://www.frontiersin.org/about/policies-and-publication-ethics#AuthorshipAuthorResponsibilities}{here} for full authorship criteria.

\section*{Funding}
The study was supported in part by the Ministry of Science and Higher Education of Russian Federation under State assignment to educational and research institutions under Project No. FSSS-2023-0009 and No. 0023-2019-0003. CHIANTI is a collaborative project involving George Mason University, the University of Michigan (USA), University of Cambridge (UK), and NASA Goddard Space Flight Center (USA).

\section*{Acknowledgments}

The authors would like to thank members of the International
Online Team “Effects of Coronal Heating/Cooling on MHD Waves” for inspiring and constructively discussing the research findings.

\section*{Data Availability Statement}
The original contributions presented in the study are included in the article, further inquiries can be directed to the corresponding author.

\bibliographystyle{Frontiers-Harvard} 
\bibliography{test}




\end{document}